# Investigation of air fluidization during intruder penetration in sand

**Bowen Wang[1,*], Yuxing Peng[2], Álvaro Vergara[1], Jordan H. Boyle[3], Raúl Fuentes[1]**

[1] Institute of Geomechanics and Underground Technology, Faculty of Civil Engineering, RWTH Aachen University, 52074 Aachen, Germany.

Bowen Wang (Corresponding author) Email: b.wang@gut.rwth-aachen.de

Álvaro Vergara, ORCID: https://orcid.org/0009-0003-4257-1442. Email: vergara@gut.rwth-aachen.de.

Raúl Fuentes, ORCID: https://orcid.org/0000-0001-8617-7381. Email: raul.fuentes@gut.rwth-aachen.de

[2] Chair of Fluid Mechanics and the Institute of Aerodynamics, Faculty of Mechanical Engineering, RWTH Aachen University, 52062 Aachen, Germany. ORCID: https://orcid.org/0009-0003-5285-3857. Email: y.peng@aia.rwth-aachen.de

[3] Faculty of Industrial Design Engineering, Delft University of Technology, 2628 CE Delft, The Netherlands. ORCID: 0000-0002-4219-9383. Email: J.H.Boyle@tudelft.nl

## Abstract

Airflow-assisted penetration offers a promising strategy for reducing resistance in granular media, with direct relevance to geotechnical applications and bio-inspired burrowing robots. However, the mechanisms governing airflow-granular interactions remain poorly understood. To address this knowledge gap, we employ a coupled computational fluid dynamics and discrete element method (CFD-DEM) approach as the primary investigative tool, validated and supplemented by experimental cone penetration tests (CPT) under varying airflow conditions. This integrated approach enables a systematic investigation of the effects of aeration on penetration resistant force. Our results reveal a nonlinear relationship between penetration resistance reduction and depth, wherein resistance approaches near-zero values up to a critical depth, beyond which well fluidization stops. Simulations demonstrate that higher airflow rates enhance the mobilization of overlying particles, thereby increasing the critical depth. A detailed meso- and micro-scale analysis of grain motion, contact forces, and pore air pressure fields further reveals four distinct penetration stages: grain ejection and channel formation, channel sealing, channel refill, and final compaction. Throughout these stages, when pore air pressure is nearly 2 times of lithostatic pressure, the penetration resistant force could drop to zero. This explicit grain-scale linkage between pore air pressure evolution and stress transfer elucidates the fundamental mechanisms behind fluidization-induced reduction in penetration resistant force, providing a quantitative framework for optimizing aeration strategies in geotechnical engineering, excavation technologies, and the design of self-burrowing robots.

* Corresponding author: B.Wang@gut.rwth-aachen.de

## List of Notations

| | |
|---|---|
| $A'$ | Projected area of particle ($m^2$) |
| $b$ | Mean magnitude of contact force (N) |
| $C_D$ | Drag coefficient (–) |
| $D_{50}$ | Median grain size (mm) |
| $D_r$ | Relative density (–) |
| e | Void ratio (–) |
| E | Particle Young's Modulus (GPa) |
| $E^*$ | Equivalent Young's modulus (GPa) |
| $\boldsymbol{F}_c$ | Sum of particle contact force (N) |
| $\boldsymbol{F}_D$ | Drag force (N) |
| $F_g$ | Gravitational force (N) |
| $F_n$ | Normal contact force (N) |
| $F_\tau$ | Tangential contact force (N) |
| $\boldsymbol{F}_{f\to p}$ | Fluid force on particle (N) |
| $\boldsymbol{F}_{p\to f}$ | Fluid momentum source term (N) |
| $\boldsymbol{F}_{\nabla p}$ | Pressure gradient force (N) |
| $G^*$ | Effective shear modulus (GPa) |
| $\boldsymbol{g}$ | Gravitational acceleration ($m\cdot s^{-2}$) |
| $\boldsymbol{I}_p$ | Particle moment of inertia ($kg\cdot m^2$) |
| $K_H$ | Hertzian normal contact stiffness ($N\cdot m^{-1}$) |
| $K_t$ | Hertzian tangential Spring stiffness ($N\cdot m^{-1}$) |
| $\boldsymbol{M}_c$ | Sum of contact moment (N·m) |
| $\boldsymbol{M}_{f\to p}$ | Moment from fluid on particle (N·m) |
| $m_p$ | Particle mass (kg) |
| $m^*$ | Equivalent mass of two contacting particles (kg) |
| $\hat{n}_z$ | Component of unit normal to the face in $z$-direction (–) |
| $n_1, n_2, n_3, n_4, n_5$ | Coarse-grain scaling factors (–) |
| $p$ | Fluid (air) pressure (Pa) |
| $p_{net}$ | net air-induced pressure (Pa) |
| $\boldsymbol{Q}_a$ | Air-probe interaction force (N) |
| $\boldsymbol{Q}_c$ | Contact resistance force on probe (N) |
| $Q_{total}$ | Total penetration resistance (N) |
| $R^*$ | Equivalent contact radius (m) |
| $Re_p$ | Particle Reynolds number (–) |
| $r$ | Polar radius in contact force distribution (N) |
| $s_\tau$ | Tangential relative displacement (m) |
| $\dot{s}_\tau$ | Tangential relative velocity ($m{\cdot}s^{-1}$) |
| $s_{\tau,max}$ | Maximum tangential displacement before sliding (m) |
| $\mathbb{T}_f$ | Fluid stress tensor (Pa) |
| $\boldsymbol{u}$ | Fluid velocity ($m{\cdot}s^{-1}$) |
| $v_l$ | Penetration velocity ($m{\cdot}s^{-1}$) |
| $\boldsymbol{v}_p, \dot{\boldsymbol{v}}$ | Particle velocity, particle acceleration ($m{\cdot}s^{-1}$, $m{\cdot}s^{-2}$) |
| $V_c$ | Volume of the computational cell ($m^3$) |
| $V_p$ | Volume of particle ($m^3$) |
| $\boldsymbol{\omega}_p$ | Angular velocity ($rad{\cdot}s^{-1}$) |
| $\alpha$ | Degree of anisotropy (–) |
| $\alpha_p$ | Void fraction occupied by particles (–) |
| $\beta$ | Preferred orientation angle (–) |
| $\delta_n$ | Normal relative displacement at contact (m) |
| $\dot{\delta}_n$ | Time derivative of contact overlap ($m{\cdot}s^{-1}$) |
| $\Delta t$ | Time interval (s) |
| $\varepsilon$ | Porosity of fluid cell (–) |
| $\zeta$ | Relative tangential displacement factor (–) |
| $\eta_H$ | Damping ratio for normal contact (–) |
| $\eta_\tau$ | Damping ratio in tangential direction (–) |
| $\theta, \varphi$ | Polar and azimuthal angles in spherical coordinate system (-) |
| $\nu$ | Poisson's ratio (–) |
| $\mu_b$ | Friction coefficient of wall (–) |
| $\mu_f$ | Gas viscosity (pa·s) |
| $\mu_p$ | Friction coefficient of particle (–) |
| $\xi_H$ | Damping coefficient ($kg{\cdot}s^{-1}$) |
| $\rho_f$ | Fluid density ($kg{\cdot}m^{-3}$) |
| $\rho_s$ | Particle Density ($kg{\cdot}m^{-3}$) |

$\nabla p$ Pressure gradient ($Pa \cdot m^{-1}$)

# 1 Introduction

Intruder penetration into soil is a long-standing and inherently complex problem in geotechnical engineering, governed by highly nonlinear interactions between the penetrating object, the granular skeleton, and the pore fluid. Among the practical realizations of this problem, the cone penetration test (CPT) represents one of the most widely used intruders in geotechnical practice. CPT is routinely applied in applications such as in-situ soil characterization (Teh and Houlsby, 1991; Robertson, 2009; Li et al., 2025), liquefaction assessment (Robertson and Wride, 1998; Youd et al., 2001; Zhang et al., 2021; Zhou et al., 2022), and foundation engineering (Meyerhof, 1951; Davidson et al., 2020; Zhang et al., 2022). Owing to its simplicity and flexibility, measurements of cone tip resistance have been extensively employed to investigate soil strength and deformation characteristics, both experimentally and numerically (Robertson, 1990, 2016; T. Zhao et al., 2023).

Despite its widespread use, intruder penetration in soils—particularly in dense sands, gravels, and highly overconsolidated materials—requires substantial force to overcome the large resistances mobilized during penetration (White and Bolton, 2005). In practice, this necessitates heavy equipment with sufficient mass to provide the required reaction force (Larsson, 1995; White, 2022). As a result, penetration becomes an energy-intensive process (Carrier, 1974; Herrick and Jones, 2002; Lunne et al., 2002; Ying et al., 2025) and its application is often limited at small or constrained sites. Additional practical challenges arise in dense urban environments, such as gravelly deposits or stiff clays, as well as in locations with limited headroom or difficult access, including sites with soft surficial soils or steep slopes (Mayne, 2007). Consequently, reducing external penetration resistance is a primary objective for facilitating soil intrusion using smaller, lighter, and more versatile equipment (Martinez et al., 2024).

To address these challenges, a variety of alternative approaches have been proposed. These include the development of advanced testing equipment (Jamiolkowski, 2012) as well as the exploration of self-burrowing technologies inspired by natural organisms, which actively modify the soil response during penetration. Such strategies include wriggling (Fudge et al., 2024), undulating (Li et al., 2023), dual-anchoring (Germann and Carbajal, 2013; Martinez et al., 2020; Tao et al., 2020; Winter et al., 2014), grabbing-pushing (Lee et al., 2020), granular fluidization (Goldman, 2006; Kamrin, 2019; MacDonald, 2015), and circumnutations and tip lateral expansion (Anilkumar and Martinez, 2024; Bengough et al., 2011). A unifying feature of these approaches is their attempt to locally alter stress transmission and particle rearrangement in the surrounding granular medium in order to reduce penetration resistance.

Among these mechanisms, granular fluidization stands out as particularly promising, as it can significantly reduce penetration resistance. Accordingly, intruders equipped with gas injection systems have been developed to promote localized fluidization near the penetrating tip (Naclerio et al., 2018, 2021). In this process, increased pore air pressure reduces the effective stress within the granular skeleton, potentially approaching near-zero values in the fluidized zone (Montana et al., 2015). Air-induced fluidization also shares mechanistic similarities with piping failure observed in certain geotechnical engineering applications in the presence of water (Richards and Reddy, 2007). Related fluidization concepts have been implemented using water as the fluidizing medium for applications such as soft soil foundation treatment (Gao et al., 2021, 2024), pile installation by jetting (Passini et al., 2018), excavation of lunar regolith (Zacny et al., 2012), and sonic drilling (Paulsen et al., 2012; Li et al., 2014; Badescu et al., 2017; Khan et al., 2025).

Despite this demonstrated potential, only limited information is currently available on the use of air-induced fluidization to reduce penetration resistance during intruder penetration (Gao et al., 2024). While experimental studies, such as those by Naclerio et al. (2018, 2021), have clearly shown substantial reductions in tip resistance through localized air injection, the underlying mechanics of this interaction remain poorly understood. In particular, critical aspects such as the real-time evolution of interparticle contact forces, the trajectories of individual particles, and the highly localized pore air pressure gradients around moving particles have not yet been resolved, either experimentally or numerically.

Hence, a numerical model incorporating two-way fluid–solid coupling at the microscale is employed in this study to enable a detailed analysis of individual particle motion, interparticle contact forces of the fluidization process and particle destabilization. In classical multiphase mixture theory, both the solid and fluid phases are typically described as interpenetrating continua using an Euler–Euler formulation (Anderson and Jackson, 1967; Drew and Passman, 1999). While such approaches are well suited for macroscopic analyses of fluidized granular media (Kaczmarek et al., 2019; Vardoulakis, 2004), they inherently homogenize particle-scale interactions and stress transfer mechanisms. In the present study, an Euler–Lagrange (CFD–DEM) framework is therefore adopted, in which the fluid phase is solved as a locally volume-averaged continuum, while the solid phase is resolved explicitly at the particle scale (Tsuji et al., 1993; Zhu and You, 2007; Sun and Xiao, 2016). This modelling strategy relies on a scale-separation assumption and enforces consistency between phases through translational interphase momentum exchange terms rather than shared constitutive laws. The approach enables direct investigation of grain-scale phenomena, such as force-chain evolution, contact loss, and localized particle rearrangements near the penetrating cone, which are central to understanding penetration resistance under air-assisted fluidization. Experimental results are used to validate a numerical investigation that allows investigating the internal synergistic mechanisms among the various components involved during the penetration process. This integrated framework allows us to resolve the drag force characteristics of particles, reveal the underlying dynamics of localized fluidization, and provide new insights into the mechanisms governing penetration resistance reduction.

# 2 Methodology

## 2.1 Governing equations for DEM

### *2.1.1 Particle motion*

In DEM, all particles are treated using a Lagrangian framework, by solving Newton's second law of motion (Cundall and Strack, 1979), that control the translation and rotation of each particle, respectively:

$$m_p \frac{d\boldsymbol{v}_p}{dt} = \boldsymbol{F}_c + \boldsymbol{F}_{f\to p} + m_p \boldsymbol{g} \tag{1}$$

$$\boldsymbol{I}_p \frac{d\boldsymbol{\omega}_p}{dt} = \boldsymbol{M}_c + \boldsymbol{M}_{f\to p} \tag{2}$$

where $m_p$ is the particle mass, $\boldsymbol{v}_p$ is the particle velocity, $\boldsymbol{g}$ is the gravitational acceleration, $\boldsymbol{I}_p$ is the particle moment of inertia, $\boldsymbol{\omega}_p$ represent angular velocity. Besides, $\boldsymbol{F}_c$ and $\boldsymbol{M}_c$ are the sum of particle contact force and contact moment, respectively. $\boldsymbol{F}_{f\to p}$ and $\boldsymbol{M}_{f\to p}$ are the fluid force acting on the particles and additional moment due to the fluid phase velocity gradient.

### *2.1.2 Contact model*

The contact force between particles is divided into normal and tangential directions. We use the Hertzian spring-dashpot model (Hertz 1881) to calculate the normal contact force:

$$F_n = K_H \delta_n^{3/2} + \xi_H \delta_{n,ij}^{1/4} \dot{\delta}_n \tag{3}$$

$$K_H = \frac{4}{3} E^* \sqrt{R^*} \tag{4}$$

$$\frac{1}{E^*} = \frac{1 - v_i^2}{E_i} + \frac{1 - v_j^2}{E_j} \tag{5}$$

where $K_H$ is the stiffness coefficient; $\delta_n$ is the normal relative displacement at the contact; $\dot{\delta}_n$ is the time derivative of the contact normal overlap, $E^*$ is the equivalent Young's modulus of the particle; $v_i$ and $v_j$ are the respective Poisson's ratios for particle *i* and particle *j*; $R^*$ is the equivalent contact radius, $\frac{1}{R^*} = \frac{1}{R_i} + \frac{1}{R_j}$ and $\xi_H$ is the damping coefficient and expressed as:

$$\xi_H = 2\eta_H \sqrt{m^* K_H} \tag{6}$$

where $m^*$ is the equivalent mass, $\frac{1}{m^*} = \frac{1}{m_i} + \frac{1}{m_j}$ and $\eta_H$ is the damping ratio for the Hertzian spring-dashpot model. The tangential force model is calculated using the Mindlin-Deresiewicz model (Mindlin, 1949; Mindlin and Deresiewicz, 1953):

$$F_\tau = -\mu F_n \left(1 - \zeta^{\frac{3}{2}}\right) \frac{s_\tau}{|s_\tau|} + \eta_\tau \sqrt{\frac{6\mu m^* F_n}{s_{\tau,max}}} \zeta^{\frac{1}{4}} \dot{s}_\tau \tag{7}$$

$$\zeta = 1 - \frac{min(|s_\tau|, s_{\tau,max})}{s_{\tau,max}} \tag{8}$$

Where $\mu$ is the friction coefficient between particles or particle to surface; $s_\tau$ is the tangential relative displacement at the contact; $\dot{s}_\tau$ is the tangential component of the relative velocity at the contact; $s_{\tau,max}$ is the maximum relative tangential displacement at which particles begin to slide; $\zeta$ is the relative tangential displacement factor; and $\eta_\tau$ is the damping ratio, which is related to the restitution coefficient of the material.

## 2.2 Governing equations for CFD

In the coupled CFD-DEM simulations, the fluid phase is regarded as a continuous phase, incompressible fluid. This assumption is justified by the maximum Mach number of 0.204 (*Ma* << 0.3, equivalent to an air velocity of ~343 m/s) of the flow in the porous media, where the fluid velocity is much lower than the speed of sound in air, making density variations negligible (Nakayama, 2018). Additionally, the CFD-DEM coupling is governed by drag and static (air) pressures as shown below and therefore, density variations are secondary since air velocities are very low. The fluid is therefore governed by the locally averaged Navier–Stokes equation (Anderson and Jackson, 1967), which is derived from the classical Navier–Stokes equations (Batchelor 2000). In this framework, the fluid-phase momentum equation follows the locally volume-averaged formulation of Anderson and Jackson (1967), in which the fluid volume fraction multiplies the pressure-gradient and viscous stress divergence terms. This reflects the fact that fluid momentum transport occurs only within the fluid-occupied portion of the representative elementary volume. The fluid phase governing equations of mean mass conservation equation, and mean momentum conservation are expressed by:

$$\frac{\partial}{\partial t}(\varepsilon \rho_f) + \nabla \cdot (\varepsilon \rho_f \boldsymbol{u}) = 0 \tag{9}$$

$$\frac{\partial}{\partial t}(\varepsilon \rho_f \boldsymbol{u}) + \nabla \cdot (\varepsilon \rho_f \boldsymbol{u}\boldsymbol{u}) = -\varepsilon \nabla p + \nabla \cdot (\varepsilon \mathbb{T}_f) + \varepsilon \rho_f \boldsymbol{g} + \boldsymbol{F}_{p\to f} \tag{10}$$

$$\mathbb{T}_f = \mu_f (\nabla \boldsymbol{u} + \nabla \boldsymbol{u}^T) + \left(\lambda_f - \frac{2}{3}\mu_f\right) \nabla \cdot \boldsymbol{u} \mathbb{I} \tag{11}$$

$$\boldsymbol{F}_{P\to f} = -\frac{\sum_{P=1}^{N} \boldsymbol{F}_{f\to p}}{V_c} \tag{12}$$

Where ε is fluid cell porosity, $\rho_f$ is the fluid density, $\boldsymbol{u}$ is the fluid phase velocity, $p$ is the fluid (air) pressure, $\mathbb{T}_f$ is the stress tensor of the fluid phase. $\lambda_f$ is the air bulk viscosity. $\boldsymbol{F}_{p\to f}$, representing the source term of momentum from interaction between particle and fluid, is the volumetric fluid-particle interaction force, calculated based on equation (12), where $V_c$ is the calculation cell volume, and $N$ is the number of particles inside the computational cell volume. $\boldsymbol{F}_{f\to p}$ accounts for the forces exerted on particle by fluid.

Porosity varies spatially and temporally during penetration due to local dilation/compaction and localized fluidization. In the present CFD–DEM framework, porosity is computed from the instantaneous particle configuration and updated at each coupling step. Consequently, porosity gradients influence the pressure and velocity fields through the evolving porosity field and through porosity-dependent closures for translational interphase momentum exchange (e.g., drag). Although alternative equivalent forms (e.g., $\nabla(\varepsilon p)$) can be written that yield an explicit $p\nabla\varepsilon$ contribution when expanded, this term is not introduced separately here to avoid double counting of porosity effects already represented by the averaged formulation and its closures (Zhu et al., 2008; Sun and Xiao, 2016). Higher-order porosity-gradient stress terms associated with specific mixture-theory formulations are not considered (Drew and Passman, 1999; Vardoulakis, 2004).

The discrete element formulation accounts for both translational and rotational motion of individual particles through explicit linear and angular momentum balance equations. In contrast, the fluid phase is modelled as a classical Newtonian continuum with a symmetric Cauchy stress tensor, and no additional balance equation for fluid micro-rotation is introduced. This corresponds to a non-polar fluid assumption. While a fully consistent treatment of angular momentum exchange would require a micropolar (Cosserat-type) fluid formulation with asymmetric stresses (Eringen, 1966; Eringen 1999; Vardoulakis, 2004), such effects are assumed to be negligible at the macroscopic scale of interest in the present study, where penetration resistance is governed primarily by drag and pressure-gradient forces.

## 2.3 Particle-fluid interaction forces

The total interaction force $\boldsymbol{F}_{f\to p}$ between fluid and particle is composed of two parts: the drag force $\boldsymbol{F}_D$ and the pressure gradient force $\boldsymbol{F}_{\nabla p}$, as follows:

$$\boldsymbol{F}_{f\to p} = \boldsymbol{F}_D + \boldsymbol{F}_{\nabla p} \tag{13}$$

The drag force is expressed according to equation:

$$\boldsymbol{F}_D = \frac{1}{2} C_D \rho_f A^{'} |\boldsymbol{u} - \boldsymbol{v}_p| (\boldsymbol{u} - \boldsymbol{v}_p) \tag{14}$$

where $A^{'}$ is the projected area of the particle onto a plane perpendicular to the direction of movement of the fluid; $\boldsymbol{u} - \boldsymbol{v}_p$ is the relative velocity between particle and fluid; $C_D$ is the drag coefficient that combines Ergun equation for higher solid density and Wen-Yu drag model for less dense flow. In particles,

$$C_D = \begin{cases} \frac{4}{3}\left(\frac{150(1-\alpha_p)}{\alpha_p Re_p} + 1.75\right); if\ \alpha_p < 0.8; \\ \frac{24 + 3.6 Re_p^{0.687}}{Re_p} \alpha_p^{-3.65}; if\ \alpha_p \geq 0.8; \end{cases} \tag{15}$$

where $\alpha_p$ is the void fraction occupied by particles (Huilin and Gidaspow, 2003); $Re_p$ is the particle Reynolds number, defined by $Re_p = \frac{\rho_f |\boldsymbol{u} - v_p| d_p}{\mu_f}$ (Ge et al.2024), where $\mu$ is hydrodynamic viscosity.

At the particle level, the force arising from a non-uniform fluid pressure field is computed by integrating the fluid pressure stress over the particle surface, resulting in a pressure-gradient force proportional to the particle volume and the local pressure gradient. This force reduces to buoyancy under hydrostatic conditions and represents a seepage-type force under spatially varying pressure fields. In the CFD–DEM framework, this term constitutes the discrete analogue of the seepage force implicitly embedded in continuum mixture-theory formulations and is balanced by the corresponding translational interphase momentum source term in the fluid equations, ensuring overall conservation of translational momentum between the fluid and particle phases. The pressure gradient force $\boldsymbol{F}_{\nabla p}$ is calculated according to the equation:

$$\boldsymbol{F}_{\nabla p} = -V_p \nabla p \tag{16}$$

where $V_p$ is the volume of particle, and $\nabla p$ is the local pressure gradient.

## 2.4 Coupling algorithm

The concept of CFD-DEM coupled analysis, pioneered by Tsuji et al. (1993), has already been used to address fluid-related geotechnical engineering and granular media problems, such as penetration into immersed granular beds (Breard et al., 2022; Lin et al., 2023; Wu et al., 2017) and particle-fluid flow in fractured formations (Ali et al., 2025; Pu et al., 2023). When the volume fraction and arrangement of particles is large enough to affect the turbulent structure of the fluid, a two-way coupling method is necessary. The particle motion is tracked in DEM, and the fluid flow can be computed in the CFD method using the Finite Volume Method (FVM). The detailed workflow of the coupling process is presented in Fig. 1. The DEM solver computes particle kinematics based on contact and hydrodynamic forces, while the CFD solver resolves the fluid flow field using locally averaged governing equations. At each coupling step, multiple DEM timesteps are performed within a single CFD time step (i.e., $\Delta t_f = N\Delta t_p$) particle information is mapped to the CFD mesh to update local porosity, and fluid forces are subsequently interpolated back to individual particles. Although CFD-DEM coupling method could solve some fluid-soil applications, fluid micro-rotation effects from particles are neglected and computational cost constraint on the total number of particles.

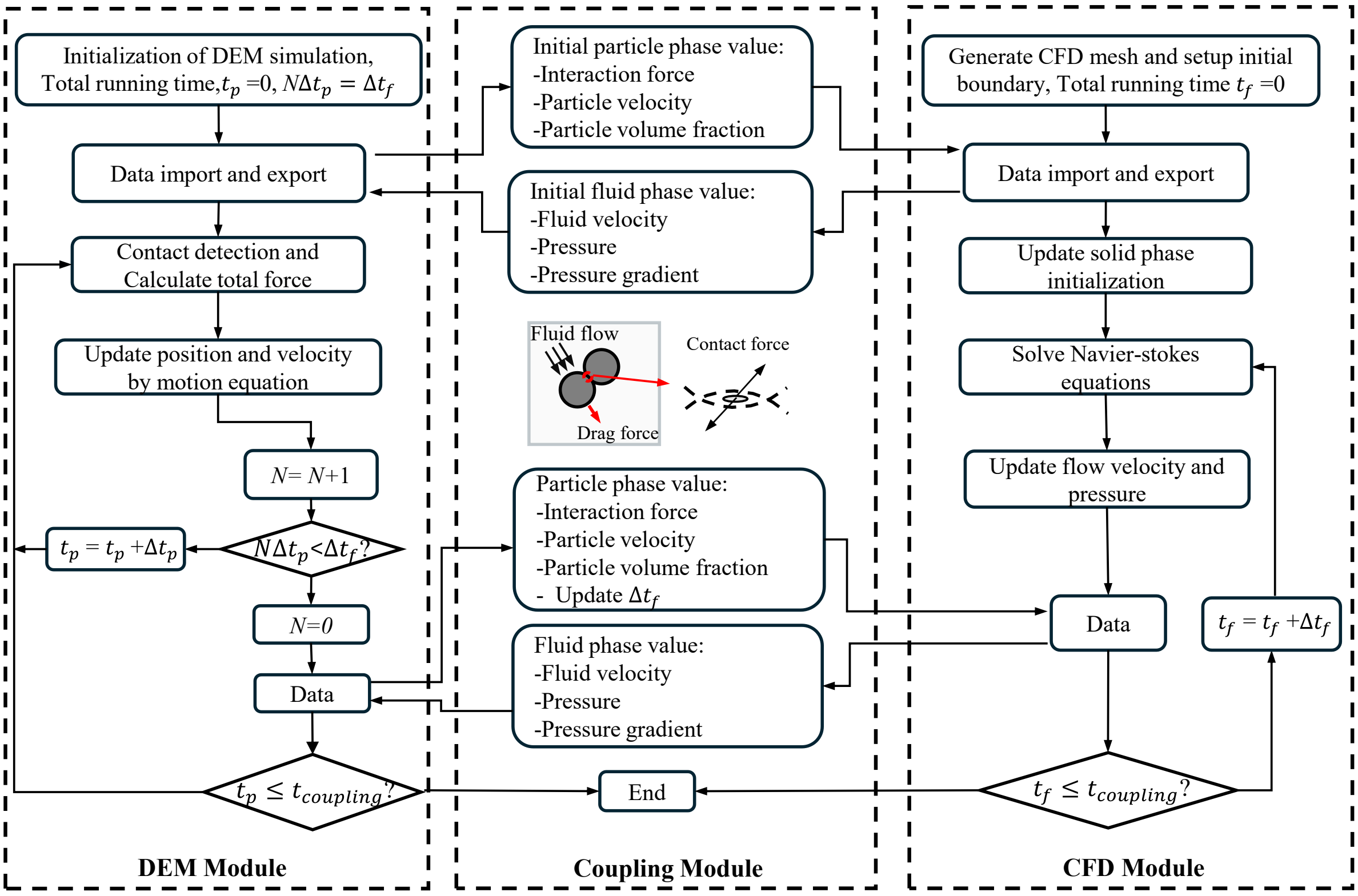

Fig. 1. Schematic diagram of CFD-DEM coupling method.

# 3 Experimental Set-ups and Numerical Model Configuration

## 3.1 Construction of the experimental study

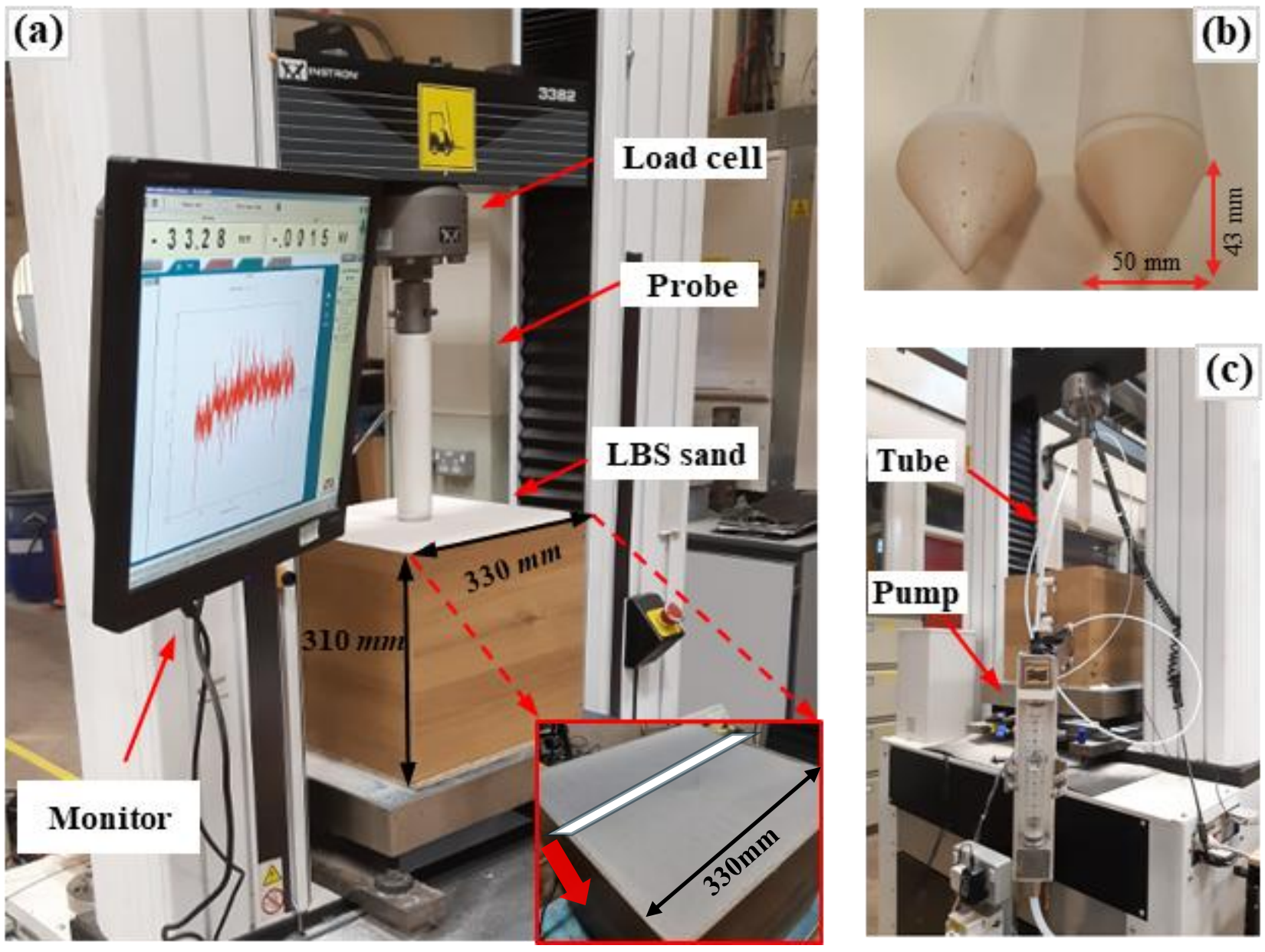

Fig. 2. Experimental set-ups and test dimensions. (a) Experimental set-ups include a loading system, comprised of a servo-controlled loading frame (American, Instron 3382) and a load cell for measuring penetration resistance, and a CPT system consisting of a 3D printing probe, an air compression pump, and a sand container with a levelled surface, as shown in the zoomed-in area; (b) Dimensions of the 3D printing probe, which features a conical tip with a diameter of 50 *mm* and an apex angle of 60°, connected to a cylinder shaft. A plastic chamber is integrated within the shaft controlled high-pressure air injection; (c) Airflow regulation system, wherein the volumetric flow rate is precisely controlled using a variable-area (VA) flowmeter (Platon GR, France).

The experimental setup comprises a wooden box, shown in Fig. 2(a), a load cell for monitoring the resistance force during penetration process, and a pump that supplies high-pressure air at a controlled flow rate (see Fig. 2(c)). The probe, with a diameter of 50 mm, was fabricated using 3D printed Acrylonitrile Butadiene Styrene (ABS). It has an apex angle of 60°, incorporating four evenly spaced holes (2.5 mm diameter), repeated in 8 equally distributed sections around the tip for controlled air injection, shown in Fig. 2(b).

The dimensions of the box are 330 mm in length, 330 mm in width, and 315 mm in height, following the container width to probe ratio of 6.6 in the experiment of Naclerio et al. (2018). The sub-angular Leighton Buzzard sand is used, with a particle size distribution ranging from 150 µm to 350 µm ($d_{50} = 0.28\,mm$). Table 1 lists its physical properties. The sand bed was prepared by gently pouring the material through a funnel from a fixed height of 10 cm above the container surface to ensure uniform deposition, avoiding any shaking or tamping. Once the container was overfilled, a straight-edge tool was used to level the surface, as illustrated in the inset of Fig. 2(a). This deposition process was repeated meticulously to obtain a medium dense sample ($D_r \approx 63\%$) (Khan et al., 2023). The experimental repeatability was ascertained using three repetitions of penetration tests under a flow rate of 10 L/min as shown in Fig. 3(a). The total penetration resistance force ($Q_{total}$) is similar in the three tests.

Table 1 Physical properties of Leighton Buzzard sand *(Cheuk et al., 2008)*

| Variable (unit) | Symbol | Value |
|---|---|---|
| Mean size (mm) | $D_{50}$ | 0.28 |
| Maximum void ratio* | $e_{max}$ | 1.010 |
| Minimum void ratio* | $e_{min}$ | 0.680 |
| Initial void ratio | $e_0$ | 0.744 |
| Specific gravity (kN/$m^3$) | $\rho_s$ | 2.65 |
| Critical state friction angle* | $\varphi_{cs}$ | 32° |

The tests were then conducted at four different airflow rates ($10, 15, 20$ and $25\ L/min$), with an additional control test at 0 $\mathrm{L/min}$ for a maximum penetration depth, $z$, of 160 mm (z/R = 6.4). The probe moved down at a constant velocity of $v_l = 5\ \mathrm{mm}/s$ once the airflow was initiated at the cone tip. To make the best record of penetration resistance, a 100 kN load cell (American, Instron 3382) with an accuracy of ±0.5% was employed, providing a resolution of $1/100$ of its full range. Airflow was supplied by a high-pressure air compressor (Atlas Copco G4FF, Sweden), and the flow rate was regulated using a variable-area (VA) flowmeter (Platon GR, France). The experimental results, presented in Fig 3(b), depict the total penetration resistance force ($Q_{total}$) recorded by the Instron machine as a function of penetration depth (up to 160 mm). The penetration resistant force shows oscillations raising from particle rearrangements during the penetration process, especially for the no-air case. As anticipated, the recorded resistance in the experiments decreases with increasing airflow rates. This reduction is primarily due to a decrease in effective stress between soil particles due to pore air pressure, in a trend consistent with previous prior experimental findings of Naclerio et al. (2018, 2021).

Particle size distribution is known to affect air–particle fluidization behaviour based on Geldart classification. Hence, we chose a narrow graded sand to isolate the coupled effects of air injection and penetration resistance while avoiding additional complexity associated with segregation, cohesive forces (fine particles) or inertial-dominated behaviour (coarse particles), which are known to alter fluidization thresholds and flow regimes (Geldart, 1973; Jackson, 2000). Our chosen sand belongs to Geldart Group B, which also includes common natural sands with small size and could be fluidized easily (Geldart, 1973). Consistent penetration trends have also been reported in previous experimental studies using sands with three times wider PSDs but still within Geldart Group B – e.g., Naclerio et al. (2021) investigated two nearly dry silica sands: Cemex Lapis Lustre 30 Mesh Sand (0.3 to 0.85 mm), and Quikrete Filter Sand (0.425-0.85mm).

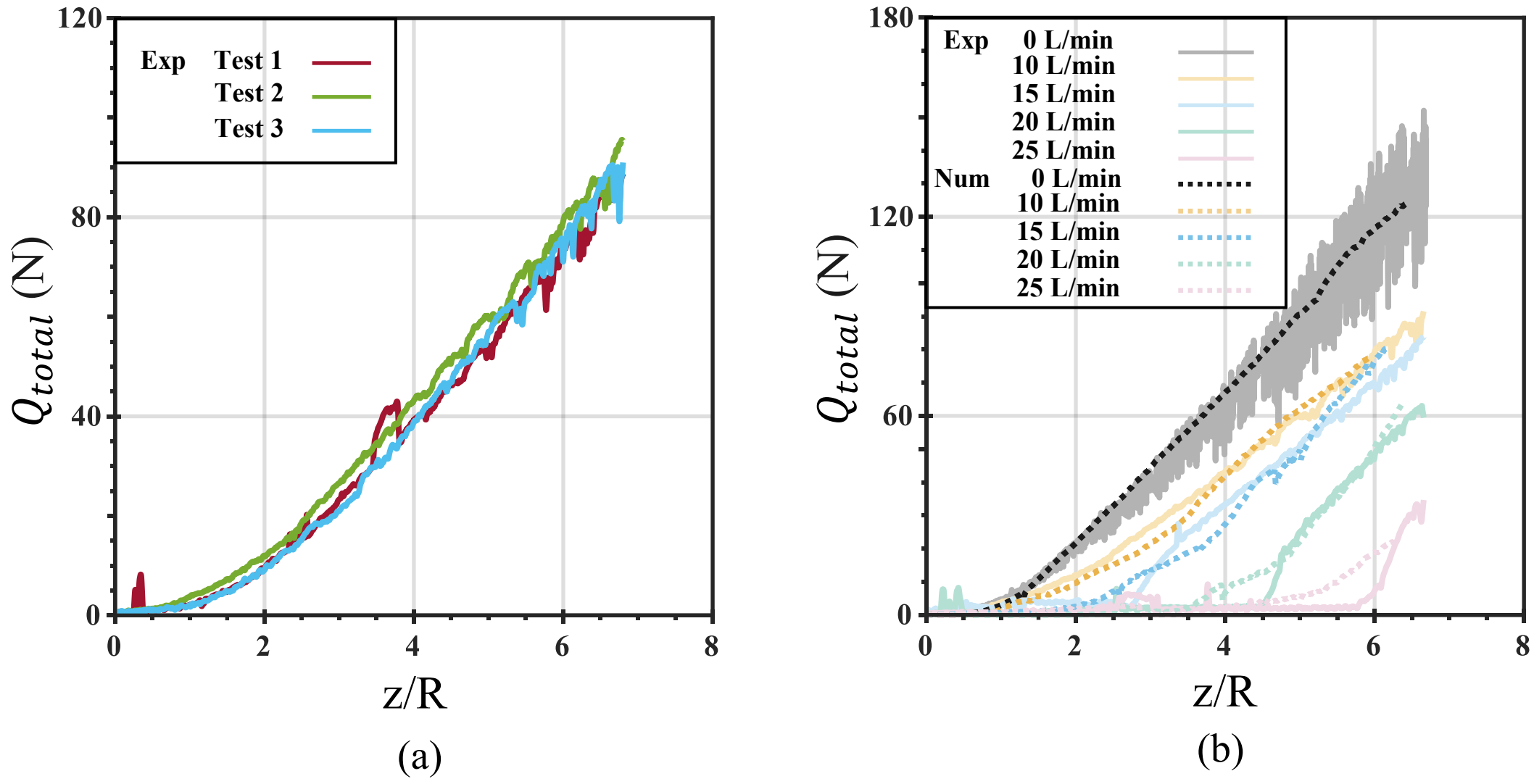

Fig. 3. Resistant force as a function of the normalized penetration depth (z/R) for a vertical penetration under various airflow rates: (a) Three repeatable results under 10 $L/min$. (b) the comparison of resistance force between experiment and numerical model under five different flow rates of 0, 10, 15, 20 and 25 $L/min$.

## 3.2 Numerical model configuration

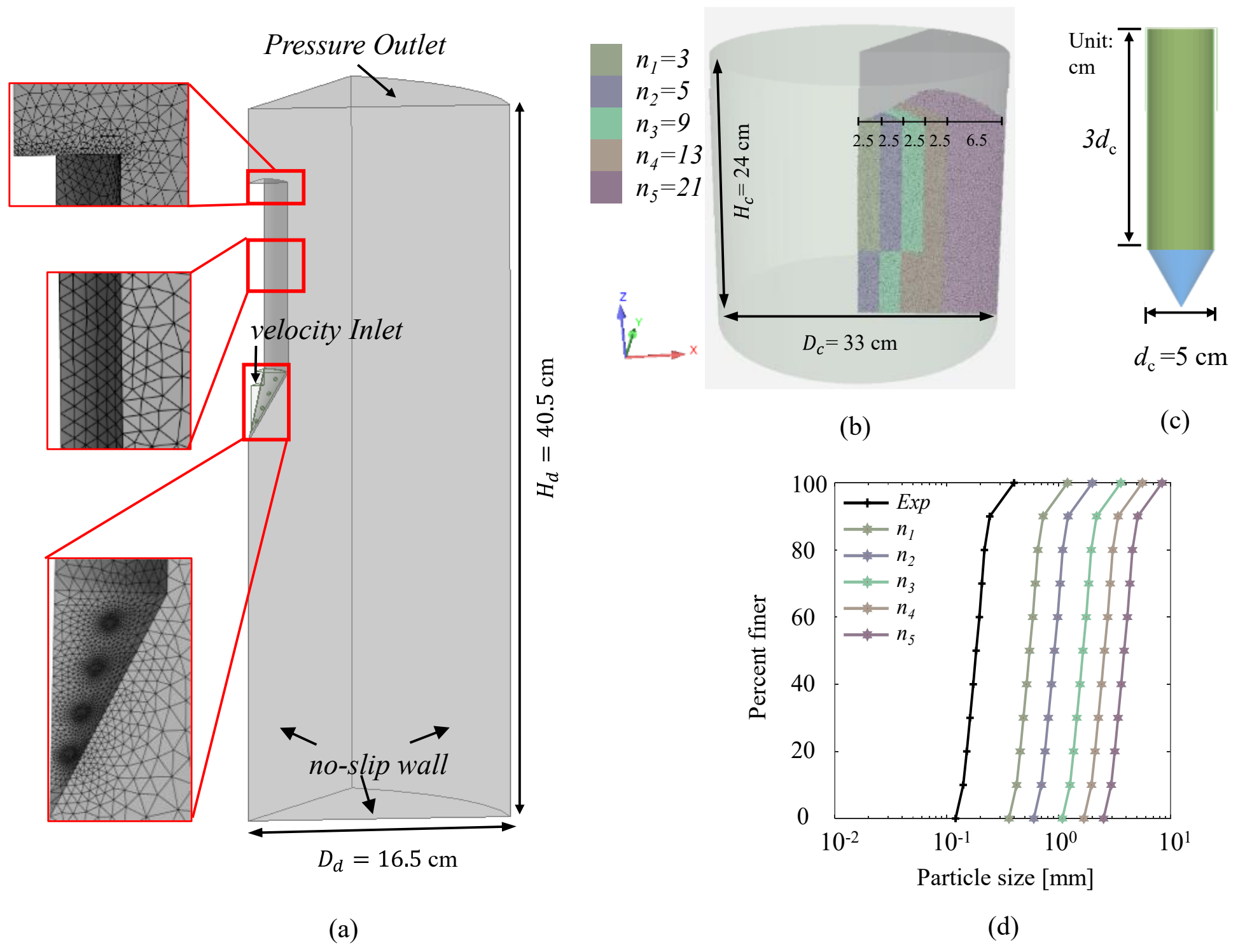

Fig. 4. CFD-DEM computational domains and particle properties used in numerical study. (a) The CFD domains, including inner airflow within the tip cone and outer airflow in the chamber, with dimensions similar to the experimental setup. The initial adaptive mesh is applied at the moving boundary of probe to ensure the high-quality mesh generation during the penetration process (not to scale). (b) view of soil chamber with multi-scale particles. (c) Geometry of the probe, with a diameter of 5 cm and a height of 15 cm. (d) Particle size distribution of Leighton Buzzard silica (LBS) sand, as utilized in both the experiment and numerical model in this study.

A coupled CFD-DEM approach is used. The computational domains comprise of a CFD part for the fluid phase in Fig. 4(a) and a DEM one for the solid phase as shown in Fig. 4(b). The fluid component of the model is implemented using Ansys® Fluent® 2024 R1 (Matsson, 2022). The detailed simulation parameters for solid and fluid phase are presented in Table 2. Because of the axisymmetry of the calibration chamber, one-sixth of the real geometry with the same dimensions as the experimental setup was modelled to limit the computational cost. This modelling strategy has also been widely adopted in previous studies, including plate load tests (Peng et al., 2023), CPT simulation with DEM (Bienen et al. 2015; Hosseini-Sadrabadi et al. 2019) and finite element method (FEM) simulations for bucket foundation installation (Zhang et al., 2024; H. Zhao et al., 2023). These studies demonstrate that such geometric simplification has a negligible impact on the accuracy or reliability of the simulation results.

The CFD domain is illustrated in Fig. 4(a). The air inlet is set at the bottom and inside the probe shaft, just above the tip, and the outlet is set at the top surface of the computational domain. Since the probe undergoes vertical motion during penetration, a user defined function (UDF) is employed to synchronize airflow with the probe movement in the DEM model. Additionally, the CFD domain is discretized using adaptive meshing to ensure high-quality mesh refinement near the moving probe at all times. In the simulation process, we defined a no-slip condition at the walls, while the airflow was governed by a velocity inlet and a pressure outlet (see Figure 4). To maintain consistency in outlet velocity between experiments and model, the airflow hole area in the numerical model is defined as one-sixth of the total hole area of the probe used in the experiment.

Fig. 4(c) show that the DEM sample creation and particle assemblage followed the process laid out in Zhang et al. (2023), resulting in an average $D_r$ of 70 % and porosity of 0.390. The minimum void ratio reported in Table 1 corresponds to a porosity of 0.400 and represents a laboratory index obtained under standardized procedures. It is used as a reference to represent the level of packing of a granular assembly, typically through relative density – in our case a dense state. In the numerical simulations, the initial particle assembly is generated by gravitational deposition and settling, resulting in an initial porosity of 0.390 (corresponding to a void ratio of approximately 0.640). This is marginally lower (~3.5%) than the porosity calculated from $e_{min}$, due to inherent uncertainties in the calculation of $e_{min}$ in the lab, as well as modelling procedures. Both represent though a similar dense state. Additionally, the model porosity agrees reasonably well (within ~7.2%) with the actual experimental value of 0.420. Such slight deviations between labs and numerically generated packing states are common in DEM due to boundary constraints (e.g., very low boundary friction coefficient of 0.05), particle discretization, and the numerical packing procedure (Shire et al., 2016; Adesina et al., 2024). Specifically, the particles size follows the distribution of the experimental sand, depicted in Fig. 4(d). In the numerical model, the probe moves at a constant low velocity of 0.015 m/s, which is three times higher than the experimental conditions. To enhance computational efficiency while maintaining an accurate representation of the sand's mechanical response, the coarse-grain model (CGM) was implemented (see details in Appendix) and its validity has been demonstrated against the experimental results. Specifically, the particle size increases with distance from the probe, with five scaling factors assigned progressively from the central penetration zone to the outer periphery: $n_1$=3, $n_2$=5, $n_3$=9, $n_4$=13, and $n_5$=21 (Chen et al., 2022; Chen and Martinez, 2023; Zhang et al., 2023). By adopting this strategy, the total number of particles was reduced to 904.320, while maintaining sufficient probe-particle interactions for reliable force measurements. The ratio of probe diameter to coarse-grained particle size is 79, ensuring adequate resolution around the probe (Zhang et al. 2023). Compared to the traditional particle refinement method (PRM) used in sample generation, the CGM approach not only reduces computational demands but also ensures accurate force interactions between particles and fluid during penetration. Experimental results of air-soil and water-soil interaction (Gao et al. 2024; Naclerio et al. 2018; Passini et al. 2018) indicate that the disturbed zone under airflow extends only up to approximately 2 times the probe radius, with particles beyond this region remaining largely undisturbed. We confirm this later in our results. In this area, CGM scaling factors of 3 and 5 were apply, remaining also well below the limit of 8 recommended for fluidized bed simulations (Di Renzo et al., 2021). In the outer zone, larger scaling factors were assigned to further reduce particle count while preserving model integrity.

A certain discrepancy in permeability between the experimental sand and the numerical sample is unavoidable due to the implementation of the coarse-grain model. In CFD–DEM simulations, permeability is not directly prescribed but rather emerges from the interplay between particle size distribution, porosity, and fluid-solid interactions. Since particle scaling increases grain size while maintaining overall porosity, the intrinsic permeability of the numerical sample becomes artificially larger than that of the real sand. Nevertheless, the present study focuses on the localized effect of airflow in the immediate vicinity of the probe. In this near-field zone, where scaling factors are kept relatively small (n = 3), the deviation in permeability remains limited and, critically, the effective particle–fluid coupling is preserved. Moreover, the disturbed zone induced by airflow does not extend beyond twice of probe radius, consistent with previous experimental observations (Gao et al., 2024; Naclerio et al., 2018). As such, the global permeability mismatch does not significantly affect the penetration resistance, which is primarily governed by local drag and contact forces around the probe. While a fully permeability-consistent CGM approach remains a challenge for future research, the present strategy provides a reasonable balance between computational tractability and accurate representation of probe–soil interaction under gas injection. However, to fully validate the effect on the overall fluid–particle interaction, expressed as the pressure drop response, we carry out a fluidised bed parametric study in the next Section.

Table 2 input parameters of CFD-DEM numerical simulation

| DEM parameters | Values | CFD parameters | Values |
|---|---|---|---|
| Particle Density, $\rho_s (\mathrm{kg/m^3})$ | 2650 | Air density, $\rho_f (\mathrm{kg/m^3})$ | 1.225 |
| Particle Young's Modulus, $E$ (GPa) | 66.5 | Air viscosity, $\mu_f$ (Pa·s) | 1.789e-5 |
| Particle Poisson's ratio | 0.45 | CFD time step, Δt (s) | 1.04e-4 |
| Friction of Particle, $\mu_p$ | 0.35 | Gravity acceleration, $g$ (m/s$^2$) | 9.8 |
| Boundary young's Modulus, $E$ (GPa) | 100 | | |
| Friction of boundary, $\mu_b$ | 0.05 | | |
| Probe young's Modulus, $E$ (GPa) | 3.8 | | |
| Friction of probe, $\mu_b$ | 0.3 | | |
| Tangential Stiffness Ratio | 1.0 | | |
| DEM time step, Δt (s) | 5.2e-7 | | |

## 3.3 Model validation

### *3.3.1 Validation against a fluidized bed model*

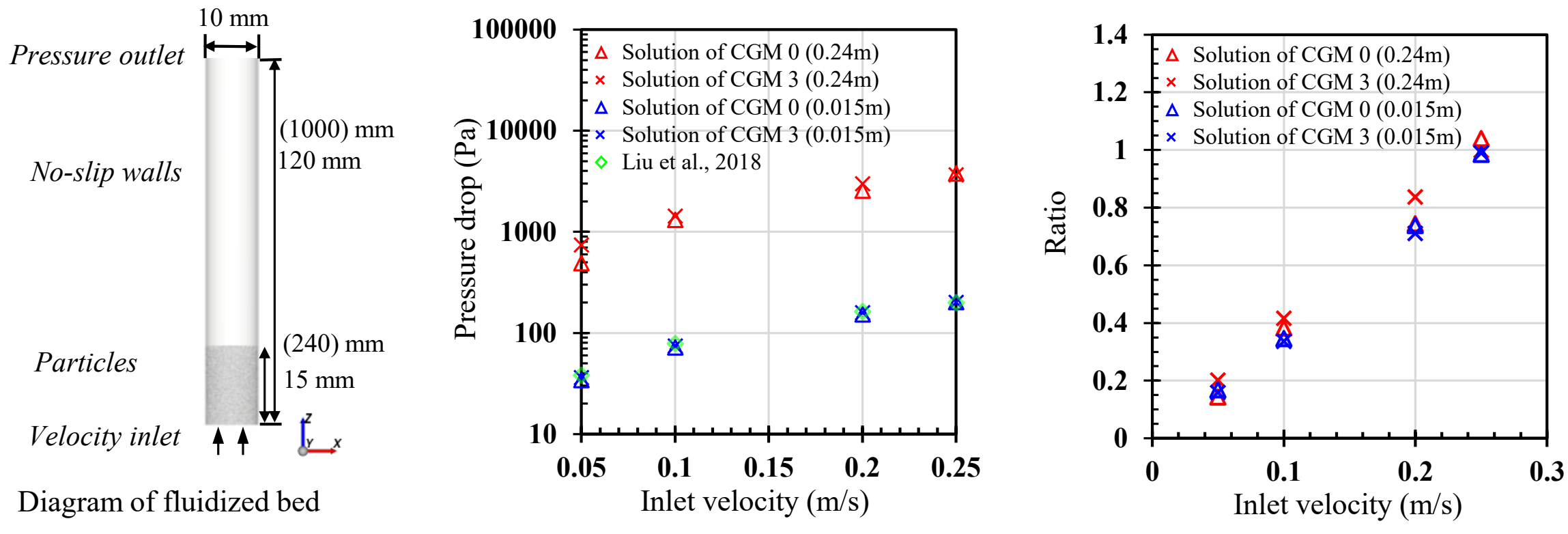

Fig. 5. (a) Diagram of the fluidized bed and (b) comparison of the predicted pressure drop between the present CGM model and the model of Liu et al., 2018. (c) the ratio of pressure drop to lithostatic pressure at the bottom of the sample.

The effect of the applied CGM particle scaling in the fluid-air force interaction is tested using a cylindrical fluidized bed case. In this case we test at original scale and a scale of 3. We also repeat the test for two bed thicknesses (15 mm and 240 mm) to understand if this affects the results. The larger thickness is close to the thickness of the bed used in our penetration experiments. The setup is presented in Fig.5(a).

The smaller model utilizing spherical particles with diameter of 0.46mm is from Liu et al, (2018), in which the minimum fluidized velocity is 0.25 m/s. We reproduce this smaller model with 15,000 particles and achieve an initial porosity of 0.408, whereas the larger model consists of 210,521 particles with an initial porosity of 0.394. Their values are very close to the 0.390 used in the numerical model of the penetration tests. This increases comparability. The bottom boundary of the cylinder was defined as a constant velocity inlet, the top boundary as a pressure outlet, and the sidewalls were assigned no-slip conditions for the gas phase. The time step in Fluent® was set as $1.18 \times 10^{-4}$ s, while the DEM time step was chosen as $2.37 \times 10^{-6}$ s for all fluidized bed models. The parameters used in this validation case are summarized in Table 3.

Table 3 input parameters for fluidized bed example

| **DEM parameters** | **Values** | **CFD parameters** | **Values** |
|---|---|---|---|
| Particle Density, $\rho_s (\mathrm{kg/m^3})$ | 2600 | Air density, $\rho_f (\mathrm{kg/m^3})$ | 1.225 |
| Particle Young's Modulus, $E$ (GPa) | 2.5 | Air viscosity, $\mu_f$ (Pa·s) | 1.789e-5 |
| Particle Poisson's ratio | 0.25 | | |
| Friction of Particle, $\mu_p$ | 0.4 | | |
| Boundary shear modulus, G (GPa) | 70 | | |
| Friction of boundary, $\mu_b$ | 0.4 | | |

As shown in Fig. 5(b), the average pressure drop between the bottom and top surfaces was evaluated across a range of inlet gas velocities. When the inlet velocity reaches 0.25 m/s, the granular bed becomes fluidized, resulting in a constant pressure drop even as the velocity becomes higher (Liu et al., 2018). The comparison in Fig. 5(b) demonstrates close agreement between the different scaled models using CGM, but critically, also with the published example. The maximum relative error not exceeding 6% and 13% for the smaller and larger model. This confirms the effect of particles on pressure drop is appropriate, validating the modelling approach. Fig. 5(c) shows the ratio of pore air pressure to lithostatic pressure is approximately 1.0 when the inlet velocity reaches 0.25 m/s. For lower inlet velocities of 0.05, 0.10, and 0.20 m/s, the ratio is 0.2, 0.4, and 0.8, respectively, showing a linear increasing trend that is expected from traditional equations for fluidised beds. When it reaches 1.0, the bed fluidises, something that happens at 0.25 m/s as already mentioned and helps confirming that the effect of the fluid on the particles is also appropriate.

### *3.3.2 Validation of current numerical and experimental model*

Secondly, we further valid the total resistance force between our experimental and numerical results, defined as $Q_{total}$. It consists of the particle-probe interaction force ($Q_c$) and air-probe interaction force ($Q_a$). The particle-probe interaction force is calculated as, $\boldsymbol{Q}_c = \sum_{i=1}^{N} \boldsymbol{F}_{z,ti} * 6$, where $\boldsymbol{F}_{z,ti}$ is the $i$-th vertical component of the force on the conical tip and shaft, and $N$ is the number of contact forces acting on the probe. The air-probe interaction force on the probe wall zone is computed by summing the dot product of the pressure and viscous forces on each face with the specified force vector. The air-probe interaction force ($Q_a$) is calculated as the vector sum of the individual force vectors for each face, $\boldsymbol{Q}_a = \sum_{i=1}^{N} p_{net} A_{probe} A \hat{\boldsymbol{n}}_{\boldsymbol{z}} * 6$, where $p_{net}$ is the net air-induced pressure, $A_{probe}$ is the probe face area, and $\hat{n}_z$ is the component of unit normal to the face in $z$-direction.

In section 3.1, the comparison between the numerical and experimental results shows strong agreement. This consistency indicates that the numerical model effectively captures the influence of airflow on penetration resistance, further validating its reliability for simulating air-assisted probe penetration testing.

# 4 Numerical results and discussion

This study aims to understand the effects of airflow with various velocities on resistant force in the granular bed. To achieve this, a meso-analysis based on the numerical and experimental result is firstly used to identify the different stages of the penetration process. Then, a micromechanical analysis, including particle displacement field and contact force is implemented to explore the probe-particle interactions. Finally, a multi-physics analysis is carried out, examining particle drag force and pore air pressure to characterize the influence of airflow.

## 4.1 Evolution of the penetration process at mesoscale

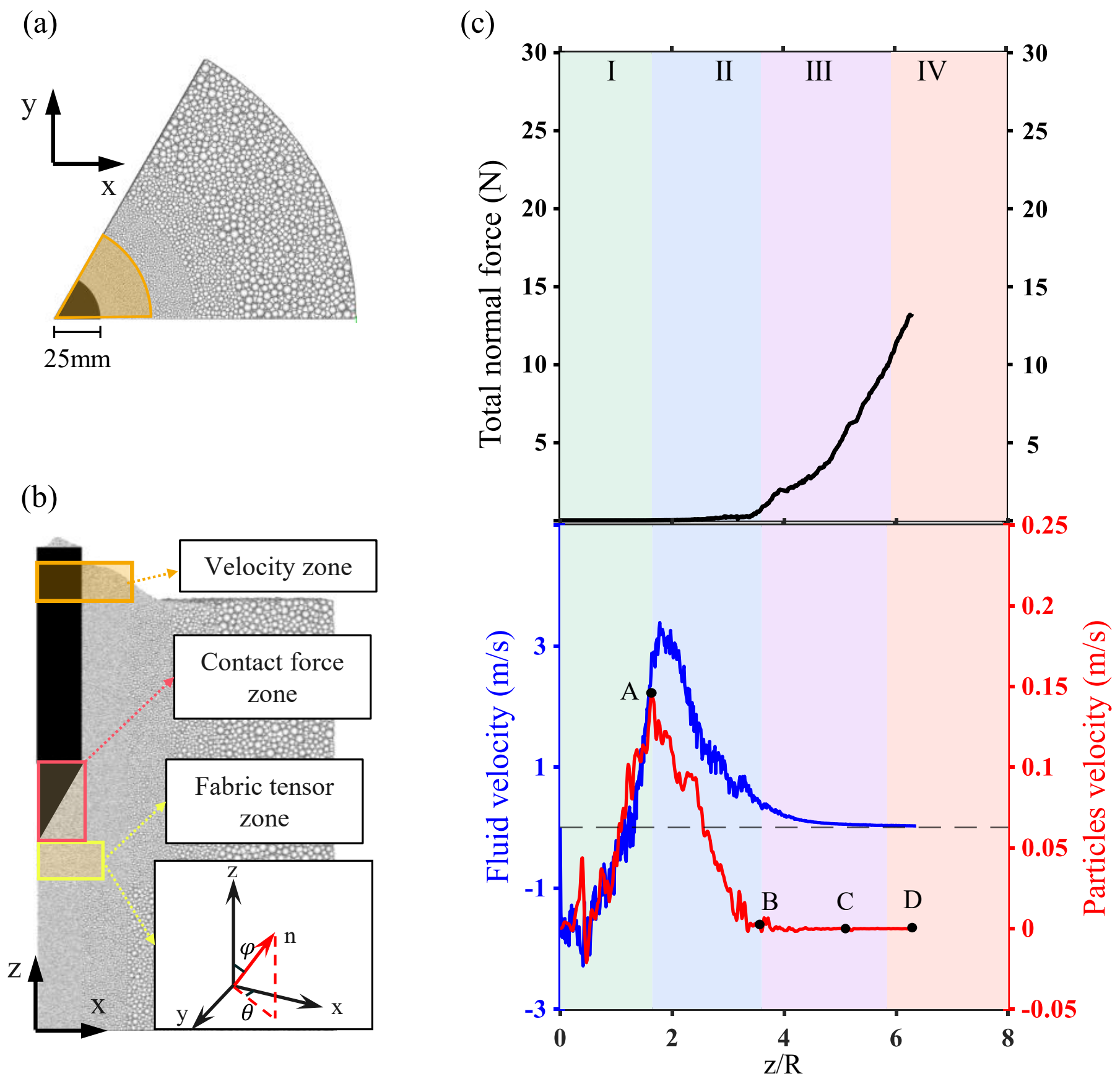

Fig. 6. The case with an airflow rate of 20 L/min is selected as a representative example to illustrate the evolution of the penetration process. (a) Plan view of selection of three representative zones around the probe for mesoscale and microscale characterization. (b) Side view of three regions, fluid velocity and particle velocity are measured in upper zone; total normal contact force between particle and intruder is measured; the region below the tip cone is used for fabric tensor and the inset figure shows three-dimensional framework for statistics of 3D contact vectors and fabric tensor analysis. (c) Characteristic variables of fluid velocity, particle average velocity along z-direction from DEM simulations, total normal force within these representative zones as a function of normalized depth.

Particle-fluid flow behaviour is affected by the particle-particle, particle-fluid flow and particle-probe interactions. The experiments clearly show an initial stage where the particles at the surface are ejected by the air outwards, moving vertically and outwards in a channel near the probe (also observed by Naclerio et al. 2018). This flow is then stopped and no further ejected particles are observed as the probe penetrates. Simultaneously, Fig. 3 already showed that the resistive force on the probe, when airflow is used, is practically null up to a point where it increases. Under the hypothesis that the observed particles movement and resistive force are linked, we define a zone at the surface where we track the particle motion using the average velocity and fluid velocity (see Fig 6(b)). Fig. 6(c) shows this evolution of average velocity of fluid and particles monitored in this zone which follow a similar trend as expected when the fluid is dragging the particles up. In combination with the total resisting force, it allows splitting the process into four Stages.

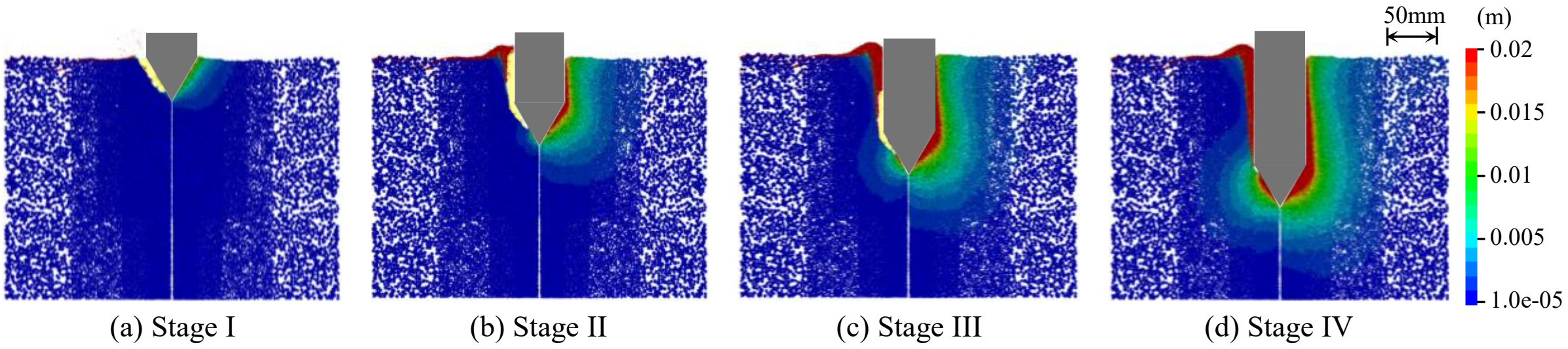

Fig. 7. Particle displacement projected on central vertical plane of sample at the end of the four stages in the case of an airflow rate of $20\ \mathrm{L/min}$ (left half) and $0\ \mathrm{L/min}$ (right half). The 'channel' between particle and probe is marked by yellow colour.

In Stage I, both particle and air velocity initially increase rapidly until the particles reach a peak value whilst maintaining an almost negligible total contact force at the probe. During this phase, the air flow ejects particles from around the probe tip, forming a crater and creating a vertical channel between the probe and surrounding particles. This channel, approximately 8.8mm wide and highlighted in yellow in Fig. 7, allows air to escape. The Reynolds number ($Re$) at the tracking region near the surface of the granular bed is about 1971 considering the channel width given above as the characteristic length. This is near the transitional stage (from laminar to turbulent). Near the probe holes, due to high velocity and small characteristic length, the flow is expected to be turbulent based on the local Reynolds number. This corresponds to the negligible resistance observed in penetration force (Fig. 3(b) and 6). Fig. 8 shows the transition value of z/R for all cases. Critically, the transition from stage I to II consistently occurs at z/R ≈ 1.8 and is independent of airflow rate. Particle ejection occurs when the local gas velocity significantly exceeds the particle terminal velocity, calculated to be ~3.48 m/s for air to drag a spherical particle of diameter equal to the simulation particles. Since all tested airflow rates (10 to 25 L/min) generate peak velocities near the nozzles that far exceed this threshold at shallow depths, channel formation always occurs.

Then, these two velocities start to slow down in Stage II until the average particle velocity becomes constant and small ($\partial v_p/\partial t \approx 0$ and $|\boldsymbol{v}_p| < 0.005\, m/s$). We chose 0.005 m/s as a value, but tested 0.01, resulting in only less than 5% difference in the transition value of z/R. At the end of this stage, the channel is sealed, and particles no longer exit. The end of Stage II also marks the moment at which the resistance force starts to raise more steeply. At this stage, the air is still flowing through the granular material but unable to drag the particles up. The terminal fluid velocity noted above, matches closely the maximum velocity shortly at the start of Stage II. Hence, the fluid is no longer capable of lifting the particles at the surface. In terms of displacement in Fig. 7(b), particles begin contacting the probe tip and start developing a small pressure bulb. This marks the onset of a gradual increase in tip resistance. As shown in Fig. 8, the transition in numerical model rises almost linearly with higher airflow rate, and deviates by less than 9% from the experimental results. The small discrepancy between the numerical model and experiments at low airflow may be due to the model's coarse-graining and limited resolution, which slightly prolongs local pore pressures and slows particle backfill. However, we consider this small discrepancy acceptable for the purpose of this work. In future work, higher-resolution simulations with smaller particles could be employed to reduce this gap below 9%. This transition indicates a critical balance between lithostatic stress and pore air pressure, which is the key factor contributing to the reduction in penetration resistance. At this transition point, the normalised pore air pressure near holes by the ratio of lithostatic pressure at the depth corresponding to the beginning of Stage III, yields values of 1.71, 1.79, 1.91, and 2.07 for flow rates of 10, 15, 20, and 25 L/min are respectively. These values closely match the normalized pore water pressure reported by Montellà et al. (2016), calculated as the ratio of local pore water pressure at the bottom of the granular bed to the corresponding effective stress of the granular skeleton. In their experiments, the granular layer is subjected to high-rate water fluid injection from below and the grain–water mixture exhibited a chimney flow regime, in which the fluidized zone extends to the free surface under high flow rates, and the normalized pore water pressure reaches values around 2.0. This behaviour is analogous to the channel formation observed in Stages I and II of our simulations. In our case the channel disappears because the material thickness increases as the probe penetrates. In fact, they demonstrated that, for a fixed injection area and bed height, there exists a critical injection flux above which the granular material above the injection zone becomes fluidized, resulting in chimney formation. In other words, the capacity of a given injection flux is limited to fluidizing a specific height of the granular bed, which is comparable to the effect of different air flow rates in our study. This insight helps explain the varying transient depths observed in our experiments under different flow rates and critically, shows that a normalised air pressure of at least 2.0 is necessary to reduce penetration resistance. The transition from stage II to stage III is governed by a balance between pore-pressure-induced reduction of effective stress and the local overburden stress; therefore, in geometrically scaled systems the transition depth is not strictly scale invariant, since increasing probe and container size increases the absolute overburden stress at a given normalized depth unless the injection conditions are scaled accordingly.

Stage III is a transitional stage, marked by a continued decrease in air velocity until it drops below 0.05m/s and stabilizes ($\partial u/\partial t \approx 0$). During this time, the resistive force increases non-linearly until the channel is fully filled by particles. Fig. 7(c) shows further development of the displacement, with displacement magnitudes decreasing radially from the probe. It is worth noting that a cavity at the cone tip is also forming at the probe tip, as also observed by Naclerio et al (2018). This is enabled by the creation of the probe shaft channel that allows particles escaping outwards. During Stage III, the channel closes and the cavity gradually collapses until it is fully refilled by particles at the onset of Stage IV. The progressive filling of this cavity reestablishes contact between the tip and the particles, directly contributing to the non-linear increase in penetration resistance observed during this stage.

In the final stage IV, the particle velocity is constant and practically null whilst a residual air velocity ranging from 0.025 m/s to 0.04 m/s applies where the air is now dissipating through the granular bed.

The permeability of the granular bed is important in both Stage III and IV, as it controls the rate of air dissipation. Lower permeability of granular bed prolongs the transitional behaviour, while higher permeability alters the streaming of air from the channel into the granular bed, accelerating channel filling and the approach to Stage IV. During this stage, the resistive force increases in a linear fashion and at a rate similar to the no-air case shown in Fig. 3. The end of Stage IV is in this case our chosen final penetration. Displacement patterns near the tip and the shaft now show downward movement due to full tip and frictional interaction, consistent with a fully developed contact regime. The linear increase in resistance with depth confirms the similarity to the no-air condition. At this stage, particle movement is mainly due to probe penetration and no air interaction.

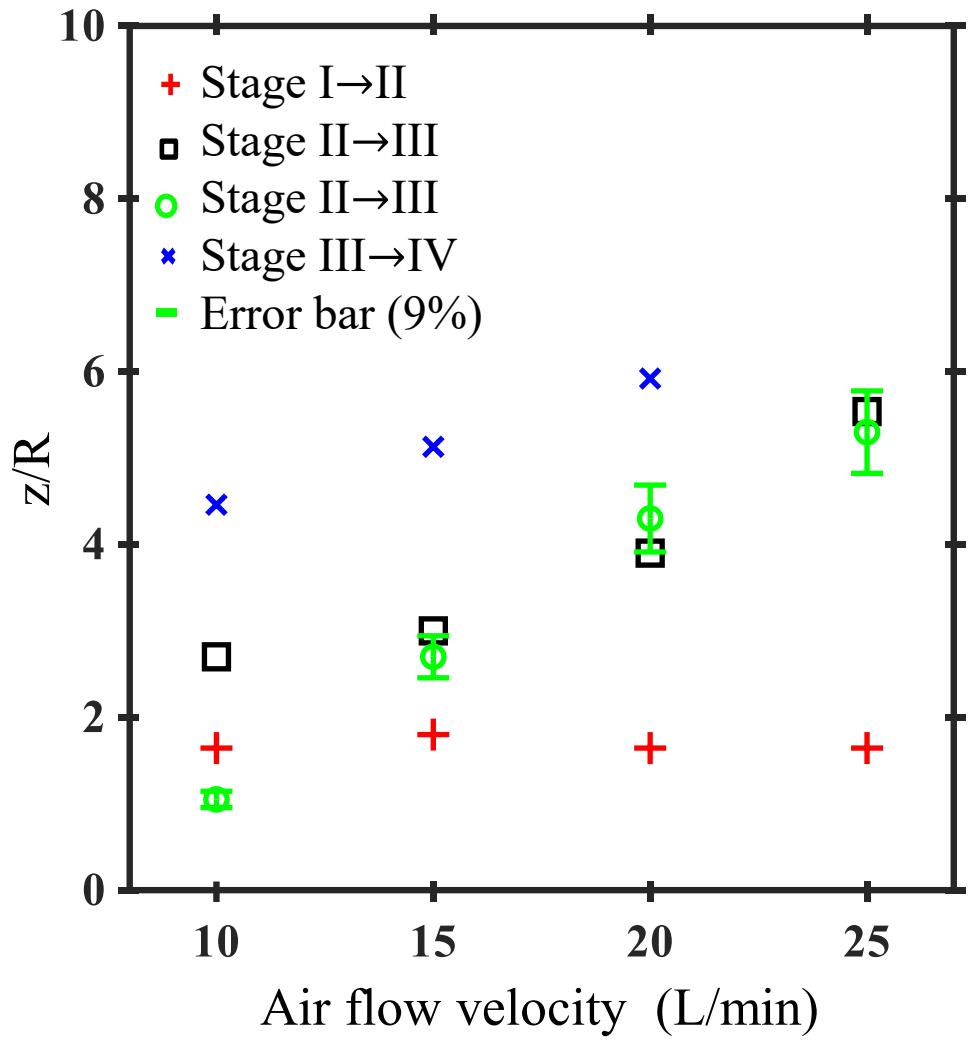

Fig. 8. The normalized penetration depth for different stage division against air flow velocity for all cases. The 9% error bar of critical depth for force transient point in experiments was marked by green.

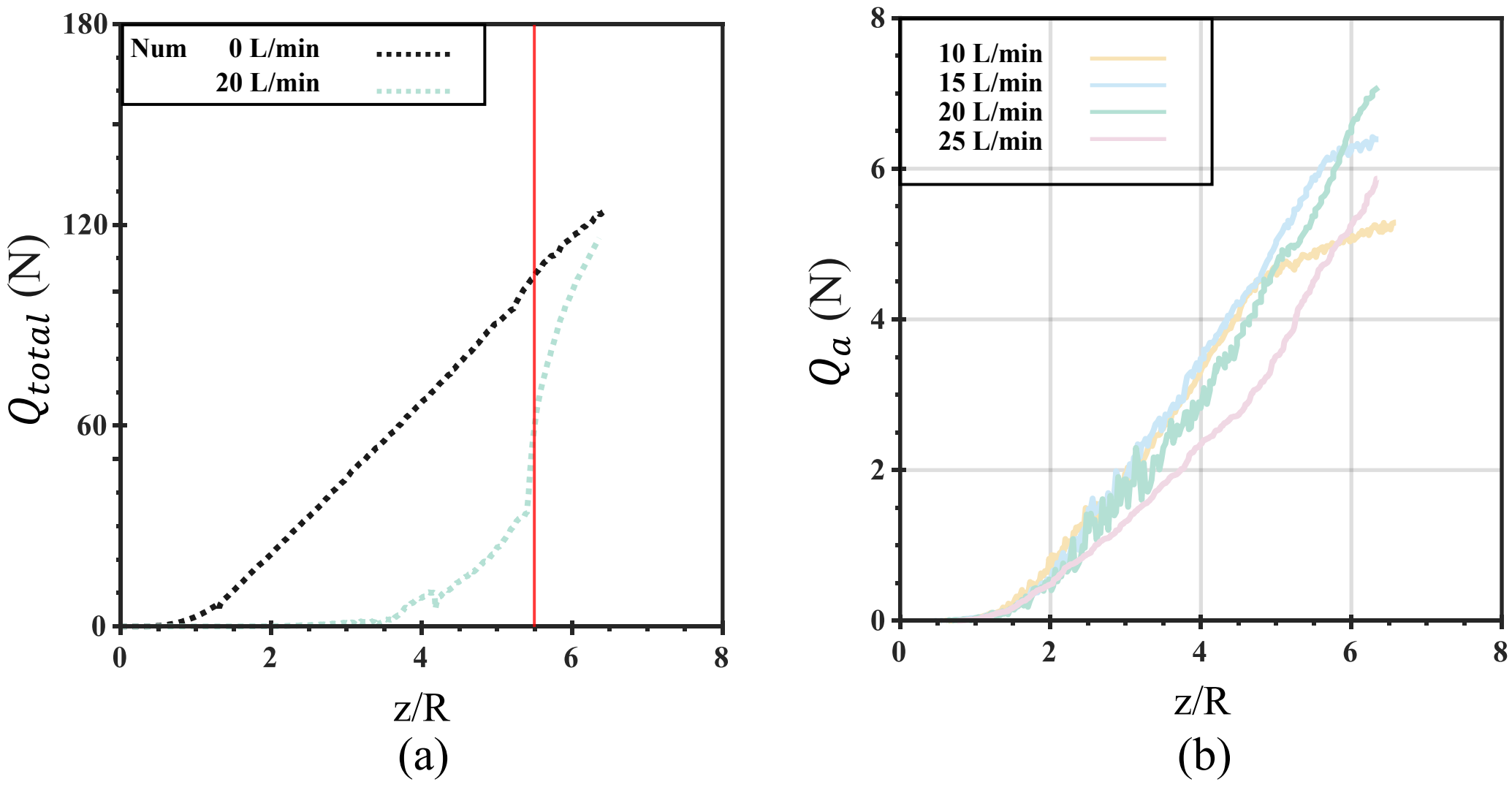

Fig. 9. (a) comparison of the resistant force without airflow (black line) and with airflow (light green line), the cease at the depth denoted by the red solid line in which the airflow is closed. (b) air-probe interaction force ($Q_a$) under four different airflow rates.

In order to prove if indeed Stage IV follows the no-air case, a numerical simulation was carried out where the high-pressure airflow was stopped shortly after the end of Stage III. The results are shown in the Fig. 9(a). It is noted that when airflow is stopped (indicated by the red solid line), the resistive force suddenly increases, reaching quickly the value that was obtained for the case without airflow. This means that air flow creates an offset to the resistive force in terms of penetration depth until the pore air pressure is ~2.0 times the effective stress of the soil. Thereafter, the resistance increases at the same rate as the no flow case.

Finally, Fig. 9(b) shows that the resistive force is overwhelmingly dominated by particle-probe interaction, as the air–probe interaction forces ($Q_a$) increase slightly with airflow rate but remain relatively small.

## 4.2 Micromechanical observations

Since the tip resistant force is mainly derived from the probe-particle and particle-particle interaction, it is necessary to explore the evolution of these interactions during the penetration process. Here, the comparison of particle displacement fields, between the traditional intruder and one with air-injection, provide more direct observation about particle movement, and the fabric tensor at the end of the intruder penetration is analysed to evaluate the contact force evolution.

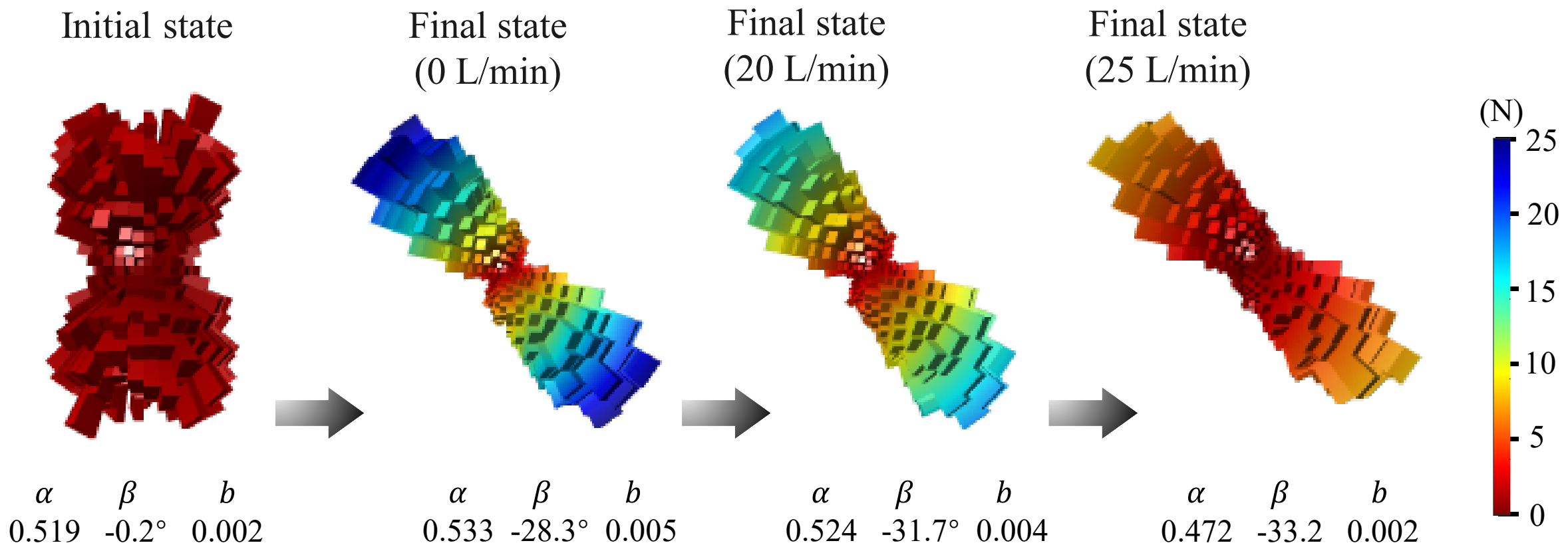

Fig. 10. Orientations of microscopic contact fabric information within fabric tensor zone shown in Fig. 5(b) under various high airflow rates. The initial state of the fabric tensor is calculated before penetration, and the final state is determined at the end of the penetration process. The height and the colour of each column in the spherical coordinate system varies with the statistical contact intensity, i.e., the higher column and warm colour means more concentrated distribution and greater contact forces. $r$ represents the polar radius, $b$ denotes the mean contact force magnitude. The degree of anisotropy is quantified by $\alpha$.

Interparticle contact force is a critical factor determining the mechanical behaviour of granular materials. The zone chosen for contact analysis is shown in the Fig. 6(b). To systematically quantify the spatial orientation of contact forces, a 3D spherical coordinate system is employed, where each contact is characterized by the polar angle $\theta$ and azimuthal angle $\varphi$, as shown in the insert of the Fig. 6(b). Each contact is mapped into the spherical coordinate system and grouped into the subregions characterized by special angular interval $(d\theta, d\varphi)$. Then, each subregion is represented by bars, lengths of which correspond to the average magnitude of contact forces within that region. Additionally, given the axisymmetry of the load about the z-axis, the influence of the polar angle $\theta$ on the evolution of the contact fabric is neglected, thereby simplifying the spatial analysis to a functional dependency on the azimuthal angle $\varphi$ (Xiong et al., 2021). The angular distribution can be further fitted using a truncated Fourier series expansion:

$$r = b\big(1 + \alpha cos(\varphi - \beta)\big) \tag{17}$$

where $r$ represents the polar radius, corresponding to the magnitude of the contact force. The parameter $b$ denotes the mean contact force magnitude, serving as an indicator of overall force intensity. The degree of anisotropy is quantified by $\alpha$. A value of $\alpha = 0$ means an isotropic force distribution, whereas larger values indicate increasing anisotropy. Finally, $\beta$ defines the preferred orientation of the anisotropic contact forces.

Fig. 10 illustrates the orientation of contact forces and the distribution of the contact forces for the initial state before penetration and final states at the end of penetration (z/R = 6.4) under three air flow conditions. Initially, the contact fabric exhibits a peanut-shape, with forces predominantly aligned in the vertical direction due to gravity-induced anisotropy. At the end of penetration, although the overall fabric shapes show no noticeable differences across all cases, a slight reduction in $\alpha$ at higher flow rates, compared to the $\alpha$ at 0 L/min case, indicates a modest decrease in anisotropy due to the perturbation of air flow. This trend is consistent with previous studies on rotational penetration (Yang et al., 2024).

Furthermore, increasing air flow velocity leads to a reduction in contact force magnitude, reflected by a modest decrease in the parameter $b$ value or warm bars, which corresponds to the overall force magnitude reduction. The reduction in contact force is associated with a local increase in pore air pressure beneath the probe, which decreases the effective stress in the sand skeleton, thereby lowering the penetration resistance. This observation aligns with the experimentally observed reduction in gas injection penetration and penetration in saturated soils (Ge et al., 2024), indicating effect of probe rotation and fluid medium on reducing penetration resistance.

## 4.3 Multi-physical analysis

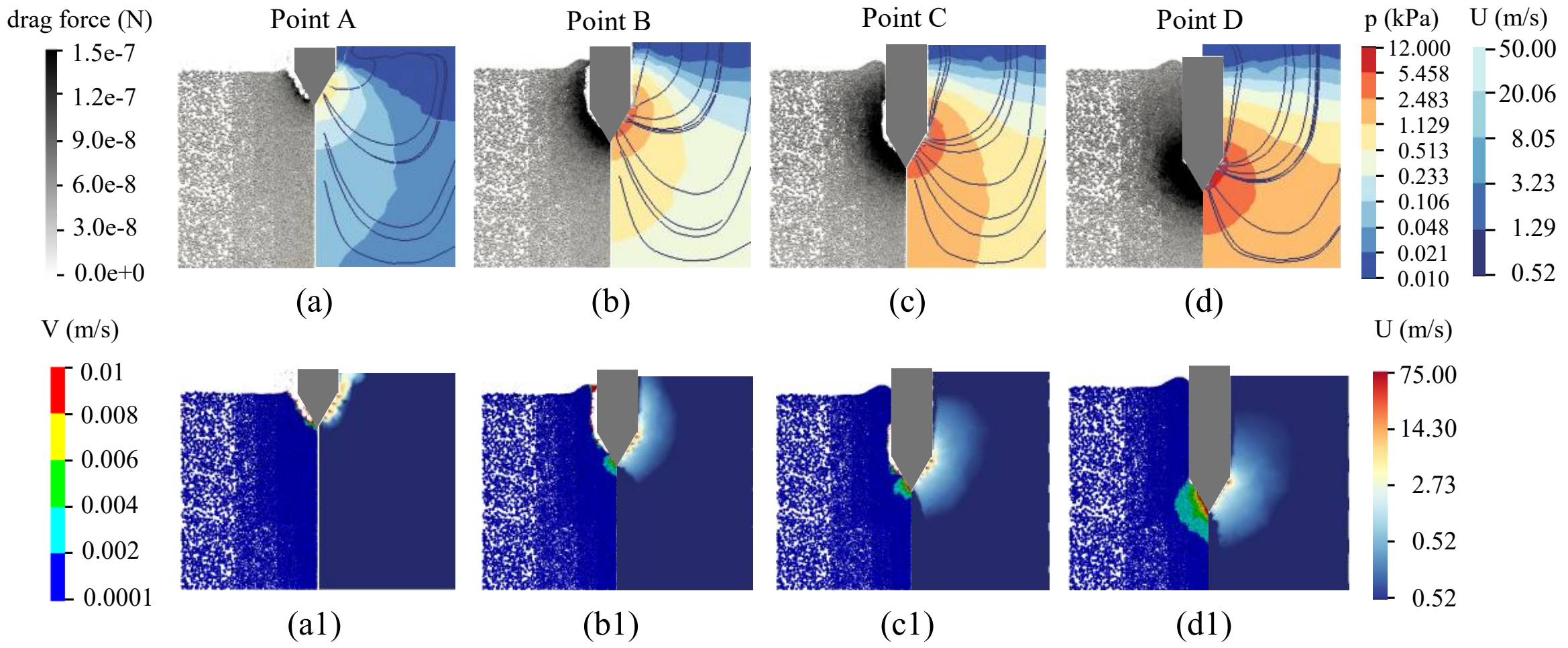

Fig. 11. (a-d) particle drag force applied on the granular bed (left half), pore air pressure and streamline in the fluid domain (right half, defined as $p$) at four characteristic points, which is shown in Fig. 6(c). (a1–d1) the velocity field including the particles (left half, particle velocity $V$) and the fluid (right half, fluid velocity $U$) in the same range. a, a1 t = 3s; b, b1 t = 6s; c, c1 t = 8s; d, d1 t = 10s.

The mechanism analysis of the air-injection penetration process involves multi-physical field evolution, characterized by interactions between airflow and granular motion. To investigate the detailed penetration process under high flow rate, the case of $20\ L/min$ has been divided into the four stages shown in Fig 6(c). At a glance, the evolution of the pressure field in Fig. 11(a)–(d) appears similar to that reported by Montellà et al. (2016). Near the injection area, the pressure contours exhibit concentric, semi-circular patterns caused by the quasi-radial flow distribution, which further supports the validity of our simulation results.

From a fluid dynamics perspective, Fig. 11(a) shows that the pore air pressure distribution during stage I is positive in the region surrounding the cone tip, while the pore air pressure in regions distant from the cone remains near zero. This indicates that the pore air pressure does not propagate to the outer layer particles. The streamline proves that air flows within channel. Particles circulating around holes experience a significant drag force from the high-pressure air flow, causing them to move upwards at high velocities, thereby creating a noticeable channel between probe and soil sample. During this phase, particle average velocity monitored in the velocity zone gradually increases, accompanied by a rise in the total particle count. The airflow velocity, shown in Fig.11 (a1), is substantial, with air exiting the probe holes and flowing directly along the channel. Furthermore, the airflow disrupts the initial particle packing configuration around the probe. This leads to fewer particle-probe interactions, keeping the resistant force at a relatively low value.

The probe reaches a critical depth at the end of stage II, from which the penetration resistance force begins to increase. Compared with the streamline distribution shown in the Fig. 11(a), it is found that the sealed channel changes the streamline. Air flow is not only in the channel but also through the surrounding particle assembly. Since the air can no longer easily escape through the sealed channel, the pore air pressure in the Fig. 11(b) around holes increase. The pressure distribution around the probe resembles that observed in cases where a granular layer is subjected to localized fluid injection (Montellà et al., 2016). Near the injection holes, the pressure contours tend to form concentric, half-circular patterns due to the quasi-radial nature of the flow. In contrast, the pressure isolines become nearly horizontal near the side walls and the surface of the granular bed. The contact force chains between particles and probe begin to form and particle-particle interactions are restored again. Subsequently, total normal force increase, see Fig. 6(c).

At the stage III, Fig. 11(c1) shows that particles no longer ejected as they did in previous stage, leading to significant decrease of effective hole area for airflow and increase of pore air pressure. The channel through which particles move becomes now gradually particle-filled, leading to a constant particle average velocity and a reduction in fluid velocity in the monitored velocity zone. In addition, the maximum velocity of particles can be always observed near the probe tip because of probe movement. The streamline has a trend through the granular bed and pore air pressure extends further away from the probe.

Fig. 11(d1) shows that the channel is fully closed at the end of stage IV and there is only a minor gap around probe shoulder due to strong airflow velocity. An obvious feature in the final stage is that streamline flows surrounding particles and Airflow directions are fully pointing outwards from the holes. In the last three stages, the pore air pressure exhibits a similar distribution, with pressure isolines (dark blue color) becoming nearly horizontal near the surface of the granular bed. This pressure distribution is consistent with the no-flux boundary condition described by Montellà et al. (2016). This indicates that no fluidization zone develops near the bed surface. Airflow velocity in the top holes reaches its maximum value, generating the highest drag force and velocity within the granular bed. Furthermore, it can be observed in Fig. 11 (b1-d1) that regions of near-zero fluid velocity appear adjacent to the upper probe shaft. This occurs because, after the vertical channel seals, the injected air can no longer easily escape upwards and is forced to dissipate radially outward into the surrounding particle assembly. The compacted granular bed above the injection zone acts as a flow restriction, creating a localized stagnation zone near the upper shaft where the fluid velocity drops to near zero.

# 5 Conclusions

This study presented a coupled CFD-DEM numerical model of air-assisted intruder penetration into granular media, validated against controlled experiments conducted under five airflow rates. Both experimental and numerical analyses were used to capture the macro- and micro-scale responses of the granular bed during penetration. The key findings are as follows:

(1) The penetration resistance force was significantly reduced as airflow increased. For example, at 25 L/min, the peak resistive force was reduced by more than 30% compared to the no-air case. This reduction is attributed to a decrease in effective stress caused by the increased pore air pressure, leading to decreased particle-particle contact forces.

(2) The penetration process is divided into four characteristic stages based on particle velocity, airflow behaviour, and resistance force evolution. The transitions between these stages were quantitatively defined using the dimensionless penetration depth z/R. These stages provide a framework to interpret the complex probe-soil-air interaction.

(3) In all cases, the pore air pressure near the probe tip needs to reach approximately twice the lithostatic stress of the surrounding soil, in order to eliminate the penetration resistant force. These findings have direct implications for probe design, suggesting what airflow rate is necessary to exceed lithostatic pressures and optimize penetration efficiency at the same time.

Further developments will explore nozzle arrangement optimization to improve airflow distribution and fluidization uniformity. Varying nozzle configurations may allow better control over drag and pore air pressure distribution, potentially reducing frictional resistance and enhancing penetration efficiency. Additionally, the conclusions of this study are based on a narrowly graded sand belonging to Geldart Group B, and extension to different particle sizes or broader particle size distributions, which are known to influence permeability and fluidization behaviour, remains an important topic for future experimental and numerical investigation.

# 6 Appendixes

## 6.1 Coarse grain model

To effectively reduce particle number, the coarse graining approach (Basson and Martinez, 2020; Bierwisch et al., 2009; Di Renzo et al., 2021; Kanjilal and Schneiderbauer, 2021) is used in this study. The coarse grain model (CGM) is developed for non-cohesive particles to represent contact, drag, and gravitational forces. The coarse grain model framework used in Ansys Rocky DEM is from an accurate treatment for Hertz-based contacts reported by Bierwisch et al. (2009). Each coarse grain particle consists of $n^3$ original particles, and its size is $n$ times larger than that of an individual original particle, where $n$ is scaling factor. The translational motion of a coarse grain particle is assumed to correspond to the average motion of the associated original particles. Therefore, the velocity and displacement of the coarse grain particle are defined as the mean values of those of the original particles.

The contact force acting on a coarse grain particle is estimated under the assumption that its kinetic energy is equivalent to the cumulative kinetic energy of the original particles it represents. During binary collisions between coarse grain particles, it is assumed that all corresponding original particles undergo binary collisions simultaneously, resulting in $n^3$ collision events. The drag force and external forces (e.g., gravitational force) are modelled by balancing the forces acting on the coarse grain particle with those acting on the corresponding group of original particles. Spherical particles are assumed, consistent with common practice in DEM–CFD simulations. Accordingly, the following relationship is established between the properties of coarse grain particles and original particles (Sakai et al., 2012, 2010):

$$\begin{aligned} m_{CGM}\dot{\boldsymbol{v}}_{CGM} &= \boldsymbol{F}_{D,CGM} - \boldsymbol{F}_{\nabla p,CGM} + \sum \boldsymbol{F}c_{CGM} + \boldsymbol{F}g_{CGM} \\ &= n^3\overline{\boldsymbol{F}}_{D,O} - n^3 V_O \nabla p + n^3 \sum \overline{\boldsymbol{F}}c_O + n^3\overline{\boldsymbol{F}}g_O \end{aligned} \tag{18}$$

where $m$, $\boldsymbol{v}$, $n$, $\boldsymbol{F}_D$, $V$, $p$, $\boldsymbol{F}_C$ *and* $\boldsymbol{F}_g$ indicate particle mass, velocity, scaling factor, drag force, particle volume, pressure, contact force and gravitational force, respectively. The subscripts of *CGM* and *O* refer to the coarse grain model and original particle. For the normal and tangential contact force, they read

$$\boldsymbol{F}c,n_{CGM} = \frac{2}{3}E^*\sqrt{R^*\delta_n}\delta_n - \xi_H\sqrt{R^*\delta_n}\dot{\delta}_n \tag{19}$$

and

$$\boldsymbol{F}c,\tau_{CGM} = -\min\left[\mu\left|\frac{2}{3}E^*\sqrt{R^*\delta_n}\delta_n - \xi_H\sqrt{R^*\delta_n}\dot{\delta}_n\right|, K_t|\delta_t|\right] sgn(\delta_t) \tag{20}$$

where $K_t$ is the tangential spring stiffness, $K_t = 8 \cdot G^*\sqrt{R^*\delta_n}$. For the drag force and pressure gradient force in CGM system, a large consensus points that grains experience the same total force as all the represented particles under the same conditions: $\boldsymbol{F}_{D,CGM} = n^3\boldsymbol{F}_{D,O}$ and $\boldsymbol{F}_{\nabla p,CGM} = n^3\boldsymbol{F}_{\nabla p,O} = n^3\, V_O\nabla p$.

## 6.2 Effectiveness of stage division for all cases

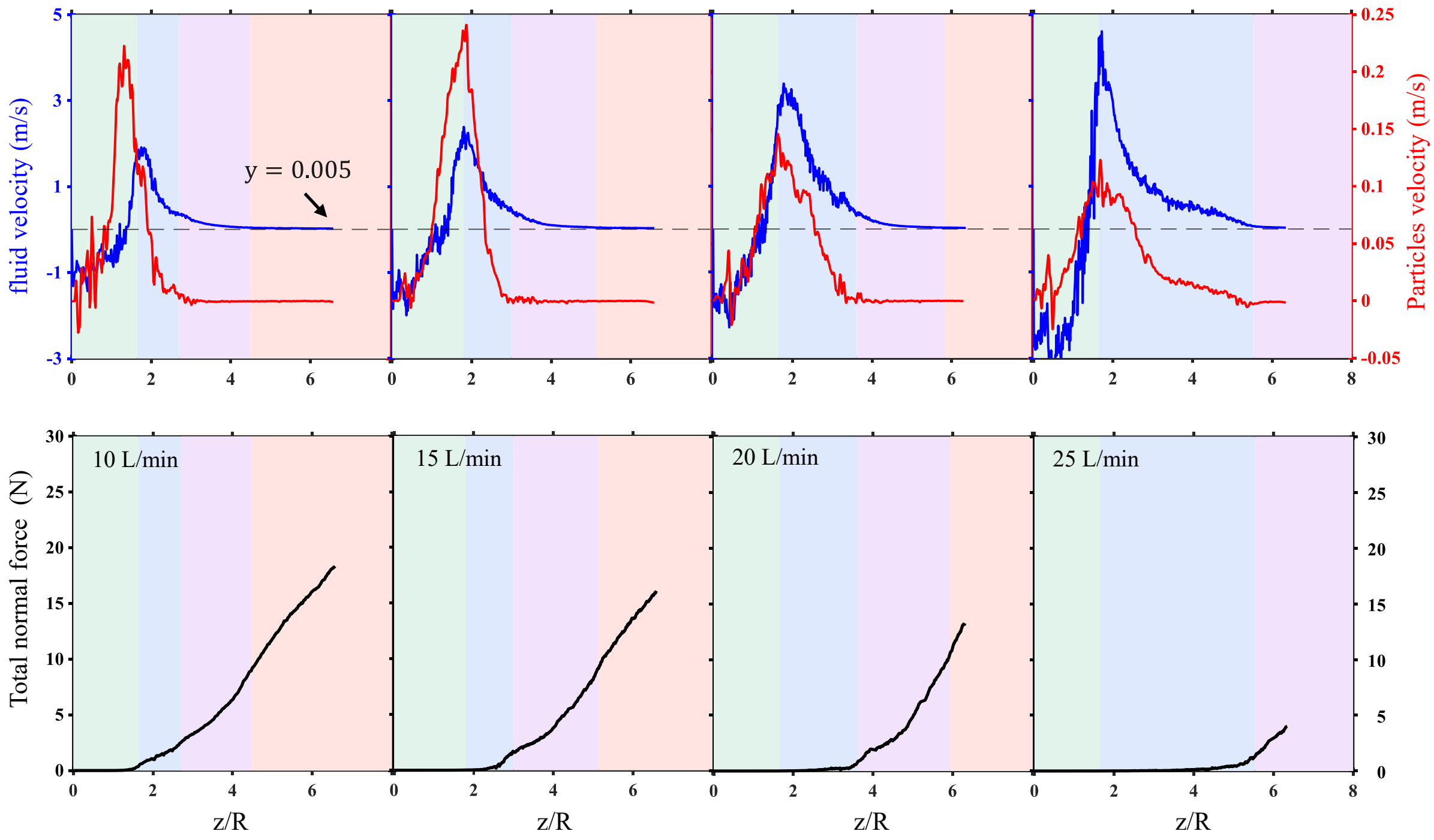

Fig. S1. The stage division from numerical model based on characteristic variables of fluid velocity, particle average velocity along z-direction and total normal force across all simulated cases.

# Acknowledgements

The first author and second author have received support from a CSC grant, which is greatly acknowledged. We thank Mr. Mingpeng Liu, Mr. Zhibin Lei, Mr. Yucheng Li at RWTH Aachen, Germany for editing the initial manuscript. A. Vergara thanks the support of the National Research Agency of Chile, grant no. 57600326, and the sponsorship of the *Deutscher Akademischer Austauschdienst* (DAAD).